\documentclass[aps,pre,preprint,superscriptaddress]{revtex4-1}

\usepackage{graphicx}
\usepackage{dcolumn}
\usepackage{bm}
\usepackage{color}
\usepackage{fancybox}

\usepackage{float}
\usepackage{epstopdf}
\usepackage{indentfirst}
\usepackage{xstring}
\usepackage{calc}
\usepackage{dcolumn}
\usepackage[dvips]{epsfig}

\usepackage[table,xcdraw]{xcolor}

\begin{document}


\title{Kolmogorov's Lagrangian similarity law newly assessed}   


%
\author{Manuel Barjona}
\author{Carlos B. da Silva}%
 \email{carlos.silva@tecnico.ulisboa.pt}
\affiliation{%
Instituto Superior T\'ecnico, Universidade de Lisboa, Av. Rovisco Pais, 1049-001 Lisboa, Portugal.\\
}%

%

\date{\today}

\begin{abstract}
Kolmogorov's similarity turbulence theory in a Lagrangian frame 
is assessed with new direct numerical simulations (DNS) of isotropic turbulence with and without hyperviscosity, which attain higher Reynolds numbers than
previously available. It is demonstrated that hyperviscous simulations can be used to accurately predict second order Lagrangian velocity structure function (LVSF-2) 
in the inertial range. The results give strong support for Kolmogorov's Lagrangian similarity assumption and allow to compute the 
universal constant of the LVSF-2, which gives $C_0=7.5 \pm 0.2$, with a new level of confidence. 
%
%
%
%
\end{abstract}

%
\maketitle
%
%
%
%

{\it Introduction}.
Turbulence arises in the motion of fluids and plasmas and is crucial for a range of diverse problems in astrophysics, 
geophysics, biology, and engineering. Almost all the existing body of knowledge on turbulence is linked to the celebrated 
{\it similarity theory} of Kolmogorov\cite{Kolmogorov-1941-LST,Monin}, 
which can predict the statistics of the velocity field $u_i(\vec{x},t)$
at fixed positions $\vec{x}$ (Eulerian frame). 

When the turbulent motion is responsible for the transport of particles a Lagrangian similarity theory is usually invoked\cite{Monin,falkovitch_rev_mod_phys_2001}, 
which is used to predict many aspects of cloud formation, combustion, pollutant dispersion and planet formation
\cite{davidson_10Chapters_book}. 
%
It is therefore surprising that, in contrast with the Eulerian similarity theory, even the most basic results from Kolmogorov's Lagrangian similarity 
theory have not yet been confirmed by either numerical simulations or experimental data~\cite{young_pof_2011}.

The key variable of interest here is the $n^{th}$-order Lagrangian velocity structure function (LVSF-n), 
\begin{equation}
D^n_L(\tau)=\overline{[\delta u_i(\tau)]^n}, 
\end{equation}
where $\delta u_i (\tau) = u_i (\vec{x}_0,t+\tau) - u_i(\vec{x}_0,t)$ is the velocity increment along a particle trajectory,
$\vec{x}_0$ is the initial particle position, $t$ is a given time instant, $\tau$ is the elapsed time, and the line '$^{\overline{~}}$'
represents an averaging operation. Statistical stationarity and isotropy imply that the probability density functions (PDF) of $\delta u_i$ ($i=1,2,3$) 
are equal and independent of $\vec{x}_0$ and $t$, and $\overline{u_i}=0$.

The Lagrangian similarity theory makes exact predictions for the LVSF-n, for time lags within an 'inertial range region' such that $\tau_\eta\ll \tau \ll \tau_L$, 
where $\tau_\eta = \left( 2 s_{ij}s_{ij} \right)^{-1/2}$ is the Kolmogorov time, and $s_{ij}=\left(\partial u_i /\partial x_j + \partial u_j/\partial x_i \right)/2$ is 
the strain rate tensor, while $\tau_L=\int\limits_0^\infty\rho(\tau)d\tau$ is a Lagrangian integral time, where 
$\rho(\tau)=\overline{u_i(t+\tau)u_i(t)} / 3 \overline{u_i^2}$ is the autocorrelation velocity function.
Specifically, it predicts that $D^n_L(\tau) \sim \tau^{\xi_n}$, where the scaling exponent is $\xi_n=n/2$.   
In particular, for the LVSF-2 self-similarity yields, 
\begin{equation}\label{eq:scaling1941}
D^2_L(\tau) = C_0 \varepsilon \tau,
\end{equation}
where $C_0$ is a universal constant, $\varepsilon=2 \nu s_{ij} s_{ij}$, is the dissipation rate, and $\nu$ is the kinematic viscosity.
This law is believed to be universal because it is linear in $\varepsilon$ and thus no intermittency corrections are required. 


Until now, and after decades of research, numerical or experimental verification of Eq. (\ref{eq:scaling1941}) 
has proven elusive. The importance of this law cannot be overemphasised, as it make the basis of virtually all the computations routinely used for turbulent
particle transport predictions~\cite{davidson_10Chapters_book}. It is generally believed that this difficulty is due to the lack of existing experimental and numerical data with sufficient accuracy at 
sufficiently high Reynolds numbers, since the existence of a range with $\tau_{\eta} \ll \tau \ll \tau_L$, strongly depends on having data at high Reynolds numbers, which is very difficult to obtain.

In the present work we carry out new (newtonian and hyperviscous) direct numerical simulations (DNS), at much higher Reynolds numbers than before, and we 
demonstrate that the hyperviscous simulations can be used to accurately predict the inertial range scaling laws of the LVSF-2. The new simulations, together with a novel analysis of the
LVSF-2 allow us to present strong new evidences in support of Kolmogorov's Lagrangian similarity theory, and to definitely establish the value of the universal constant $C_0$.


~\\
{\it Direct numerical simulations}. 
Several DNS of statistically stationary (forced) isotropic turbulence in a periodic box with sizes $2 \pi$ including point particles (tracers), 
were carried out using a classical pseudo-spectral code, previoulsy used in \cite{VSP2014,valente_pof_2016}, 
to numerically integrate the hyperviscous Navier-Stokes equations~\cite{borue_EPL_95,borue_jfm_98,pope_pof_05_hyperviscous},
%
\begin{equation}\label{eq:momentumHVISC}
\frac{\partial  u_i}{\partial t}+u_j\frac{\partial u_i}{\partial x_j}= -\frac{1}{\rho}\frac{\partial p}{\partial x_i}+ (-1)^{h+1}\nu_h\Delta^h u_i+ f_i,
\end{equation}
where $u_i$ and $p$ are the velocity and pressure fields, respectively, while $f_i$ is an artificial forcing, which is uncorrelated with the velocity field
and delta-correlated in time\cite{VSP2014}. In all the simulations the total power input forcing $P$, which on average equals the viscous dissipation rate 
$P=\varepsilon$, is equal to $P=10$ (m$^2$s$^{-3}$), and the forcing is imposed on the first $2$ wavenumbers, and is concentrated in wavenumber $k_f=2$.
$h$ is the order of the hyperviscosity and $\nu_h$ is the corresponding hyperviscosity ($\rho$ is the fluid density).
The Navier-Stokes equations are recovered for $h=1$, while $h \ne 1$ corresponds to the hyperviscous simulations.  
Table \ref{table_DNS} summarises the DNS used in this work. 
The number of tracked particles $N_p$ increases with $N$, attaining $N_p=1,2$ million tracers for the biggest DNS. 
The particles tracking uses the same (3rd-order) Runge-Kutta time-stepping scheme used in the Eulerian DNS~\cite{VSP2014},
and a cubic interpolation is used to interpolate the velocity into the particle positions. Full details are given in ~\cite{barjona_msc}. 

 \begin{table}[!htb]
  \caption{
  Parameters of the DNS without ($h=1$) and with ($h=8$) hyperviscosity 
  (left and right sides of the table, respectively):
  %
  Number of grid points ($N^3$);
  Reynolds number based on the Taylor micro-scale ($Re_{\lambda}$);
  Kinematic viscosity ($\nu$); 
  Ratio between the integral and Kolmogorov time scales ($\tau_L/\tau_{\eta}$);
  Resolution ($k_{max} \eta$);
  Wavenumber corresponding to the maximum enstrophy in the hyperviscous simulations ($k_d$);
  %
    Location of the peak maximum of $D^2_L(\tau)$  ($\tau_0^{*}$);
    Maximum of $D^2_L(\tau)/(\varepsilon \tau)$ ($C_0^{*}$);
    Location of the inertial range peak maximum of $\zeta_2'(\tau)$ ($C_0^{* *}$);
   $\zeta_2(\tau)$ for $\tau=\tau_0^{* *}$ ($\alpha$);
   Universal constant of the LVSL-2 computed through Eq. (\ref{eq:c0**}) ($C_0^{* *}$). 
   }
   \label{table_DNS}
  \begin{center}
    \begin{tabular}{lccccccccccclcccccccc}
     \hline
                    $N^3$    & $Re_{\lambda}$ &  $\nu$ & $\tau_L/\tau_{\eta}$ & $k_{max} \eta$ & $\tau_0^{*}/\tau_{\eta}$ & $C_0^{*}$ & $\tau_0^{* *}/\tau_{\eta}$ & $\alpha$ & $C_0^{* *}$ &  &  &
                    $N^3$    & $Re_{\lambda}$ &  $\tau_L/\tau_{\eta}$ & $k_d$ & $\tau_0^{*}/\tau_{\eta}$ & $C_0^{*}$ & $\tau_0^{* *}/\tau_{\eta}$ & $\alpha$ & $C_0^{* *}$ \\
      \hline       
                    $32^3$    &  $24$   &   $0.1$       & $3.4$   & $1.6$  & $3.7$ & $2.0$ & $-$  & $-$ & $-$ & &  &  $128^3$    &  $276$   &   $13.5$  & $24$   & $4.0$ & $5.2$ & $-$ & $-$ & $-$ \\
                    $64^3$    &  $50$   &   $0.04$     & $5.2$   & $1.6$  & $3.8$ & $3.0$ & $-$  & $-$ & $-$ & &  &  $256^3$    &  $450$   &   $22.1$   & $54$  & $4.3$ & $5.7$ & $9.0$ & $0.82$ & $7.8$ \\
                    $128^3$  &  $88$   &   $0.015$   & $7.7$   & $1.5$  & $4.0$ & $4.0$ & $-$  & $-$ & $-$ &  &  &  $512^3$    &  $701$   &   $35.9$  & $105$ & $4.4$ & $6.0$ & $9.0$ & $0.88$ & $7.4$ \\
                    $256^3$  &  $131$ &   $0.0071$ & $12.2$  & $1.8$  & $4.2$ & $4.4$ & $-$ & $-$ & $-$ &  &  &  $1024^3$  &  $1102$  &   $49.1$  & $207$ & $4.7$ & $6.3$ & $9.0$ & $0.91$ & $7.4$ \\
                    $512^3$  &  $228$ &    $0.0025$  & $17.7$  & $1.6$  & $4.4 $ & $5.1$ & $10.5$ & $0.78$ & $7.7$ &  &  &  $2048^3$  & $1744$ &  $88.7$ & $412$ & $4.9$ & $6.6$  & $9.0$ & $0.9$2 & $7.6$ \\
                    $1024^3$  &  $381$ &  $0.001$    & $30.8$   & $1.6$ & $5.2$ & $5.7$ & $10.5$ & $0.86$ & $7.4$ &    &   &   &    &        &  & & & \\
      \hline
    \end{tabular}
  \end{center}
\end{table}

A set of $6$ Navier-Stokes DNS ($h=1$) were carried out with Reynolds numbers of up to $Re_{\lambda}=381$ and resolutions of $k_{max}\eta \approx 1.6$,
essentially to demonstrate that hyperviscosity does not affect the Lagrangian statistics in the inertial range. 
A total of $5$ hiperviscous DNS was carried out with $h=8$, and a hyperviscosity obeying the relation $\nu_h \left( N/2 \right)^{2h} \Delta t \approx 0.5$, 
where $\Delta t$ is the time step of the simulations~\cite{borue_EPL_95,borue_jfm_98}. 
The Reynolds numbers of the hyperviscous DNS is given by 
$Re_{\lambda}=C_8 (k_d / k_f)^\frac{2}{3}$, where $C_8=50$ and $k_d$ is the peak enstrophy wavenumber. 


~\\
{\it Lagrangian statistics from hyperviscous simulations}.
By concentrating the viscous dissipation on a small range of high wavenumbers near the maximum $k_{max}$, hyperviscous simulations
substantially increase the extent of the inertial range region compared to 'Newtonian' ($h=1$) DNS, which allows to attain much higher 
Reynolds numbers~\cite{borue_EPL_95,borue_jfm_98,pope_pof_05_hyperviscous}. Recently, hyperviscous simulations were used to study the shape of the 
energy spectrum in viscoelastic turbulence \cite{valente_pof_2016}, and here we show that this technique can be used to study in detail the 
Lagrangian statistics of turbulence for inertial times $\tau_{\eta} \ll t \ll \tau_L$. The realisation that hyperviscosity can be used to study the Lagrangian
statistics in the inertial range is an innovative aspect of the present work, which should not be surprising. Recall that in virtually all similar DNS studies 
the large scales are also forced, and thus are not an exact solution of the Navier-Stokes equations, however this does not prevent the study of turbulence statistics in the inertial range.

\begin{figure}[h] 
\centering
\includegraphics[width=0.75\textwidth]{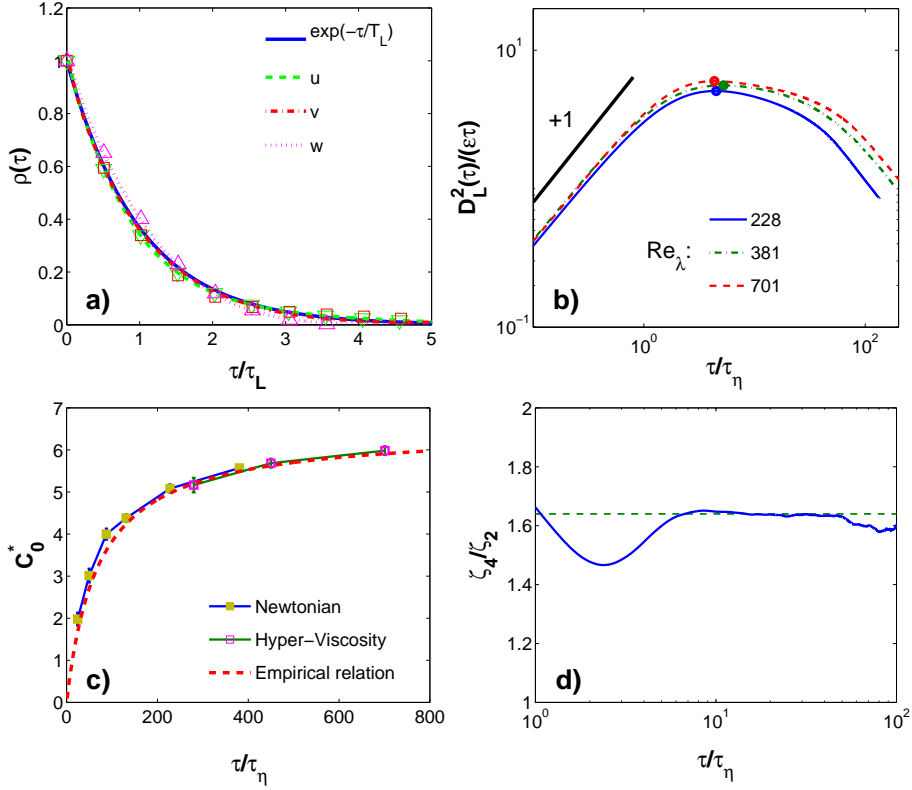}
\caption{
(a) Lagrangian correlation function for $u_i=(u,v,w)$ for the Newtonian (symbols) and hyperviscous (lines) DNS with $N^3=256^3$.
The function $e^{-\frac{\tau_L}{T_L}}$ is also added for comparison;
(b) Normalised LVSF-2 for some Newtonian and Hyperviscous simulations listed in table \ref{table_DNS} at increasing Reynolds numbers; 
(c) Evolution of the peak value $C_0$ defined in Eq. (\ref{eq:scaling1941}) as function of Reynolds number, for several Newtonian and hyperviscous simulations,
compared with the empirical relation $C_0^{\ast} = 6.5 / \left( 1 + 70 / Re_{\lambda} \right)$, from \cite{Sawford_dispersion}.
(d) Scaling coefficient of the LVSF-4 ($\zeta_4$) obtained from the hyperviscous simulation with $N^3=1024^3$ ($Re_\lambda=1102$). 
The green dashed line has the constant value of 1.66.
}
\label{hyper_validation}
\end{figure}

Table \ref{table_DNS} shows that the hyperviscous DNS with $2048^3$ grid points attains a Reynolds number of $Re_{\lambda} \approx 1700$, which is much 
higher than in previous numerical works~\cite{Sawford_dispersion,young_pof_2011}, and, as we will see below, allows for the first time to directly observe the Lagrangian Kolmogorov similarity. 
Specifically, extensive validation tests have shown that for $\tau_{\eta} \ll t \ll \tau_L$, Lagrangian statistics from Navier-Stokes ($h=1$) and hyperviscous ($h=8$)
simulations at the same Reynolds number are virtually equal. Figures \ref{hyper_validation} (a-d) display several of these results while other tests 
are described in \cite{barjona_msc}. 

Figure \ref{hyper_validation} (a) shows the Lagrangian correlation function $\rho(\tau)$ for the velocity components $(u,v,w)$ in the Newtonian and hyperviscous 
simulations corresponding to $N^3=256^3$. The agreement between the Newtonian and hyperviscous results is very good, and moreover all the correlations
exhibit a clear exponential decay as predicted in \cite{yeung_pope_jcp_82}. Furthermore, $\rho(\tau)$ obtained for $u$, $v$, and $w$ is very similar which shows 
that the forcing does not impose any significative level of anisotropy in the present simulations. 

\begin{figure}[h]
\centering
\includegraphics[width=0.75\textwidth]{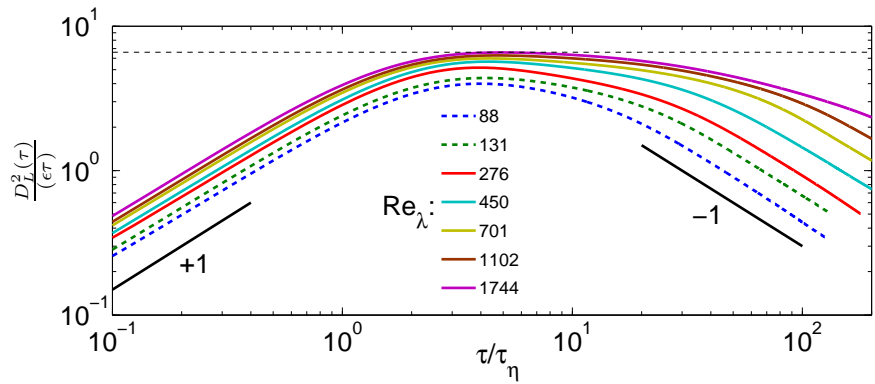}
\caption{
Lagrangian velocity structure function of order 2 (LVSF-2), normalised by $\varepsilon \tau$, 
as function of the time lag $\tau$ for several Reynolds numbers from the Newtonian and 
hyperviscous simulations (listed in table \ref{table_DNS}). The horizontal dashed line is at $C_0=6.6$. 
}
\label{fig:PLOTLVSF_HVISC}
\end{figure}

Figure \ref{hyper_validation} (b) shows $D^2_L(\tau)$ normalised by $\varepsilon \tau$, obtained with Newtonian and hyperviscous simulations, 
for increasing Reynolds numbers. First, in both cases a slope of $+1$ is obtained in the interval $\tau<\tau_{\eta}$ as expected \cite{yeung_pope_jcp_82}
(see also Fig. \ref{fig:PLOTLVSF_HVISC}).
Secondly, for $\tau>\tau_L$ a slope of $-1$ is recovered, again as expected since $\rho(\tau)$ vanishes. Thirdly, the peak value ($C_0^{\ast}$)  
increases with the Reynolds number, regardless of whether the simulations are Newtonian or hyperviscous. Finally, the exact location of the peaks ($\tau_0^{\ast}$) also
shows the consistency of the hyperviscous results {\it i.e.} the Newtonian simulations this peak occurs at a time lag $\tau_0^{\ast}$, which is slightly higher 
than in the hyperviscous simulations, but since this peak is in the transition between the dissipative and integral time scales the slight smaller location of the peak
in the hyperviscous simulation is actually consistent with the decrease of the width of the dissipative length scales in these simulations.

The definitive demonstration that hyperviscous simulations can accurately predict the Lagrangian statistics at inertial time lags is shown in the next two figures.
Figure \ref{hyper_validation} (c) shows the constant $C_0$ defined in Eq. (\ref{eq:scaling1941}) for the Newtonian and hyperviscous simulations,
as function of Reynolds number, for $Re_{\lambda} \le 800$, together with the empirical relation $C_0^{\ast} = 6.5 / \left( 1 + 70 / Re_{\lambda} \right)$
from \cite{Sawford_dispersion}, which is used here only to compare the present results with $C_0$ obtained in previous numerical simulations.
It is clear that $C_0$ computed from the Newtonian and hyperviscous simulations are virtually equal for the same Reynolds number. Moreover, the 
presents values of $C_0$ have excellent agreement with previous numerical simulations. 

Finally, figure \ref{hyper_validation} (d) shows the scaling coefficient $\zeta_4/\zeta_2$ from the extended similarity theory\cite{benzi_jfm_2010}, where
$\zeta_n = d [ \log D_L^n (\tau) ] / d [ \log D_L^2 (\tau) ]$, for the hyperviscous simulation with $N^3=1024^3$ ($Re_\lambda=1102$), 
as function of the time lag (note that $\zeta_2=1$). Recall that the this expression allows the computation of inertial scaling coefficients even in the absence of an extensive inertia range.
$\zeta_4$ was averaged in the interval $10 \le \tau /\tau_{\eta} \le 40$, and the uncertainty estimate of the scaling coefficient uses the maximum difference between 
$\zeta_4$ computed with the mean value of the $(u,v,w)$ velocity components, and $\zeta_4^i$ computed using only the i-th velocity component. 
For inertial range times $\zeta_4$ is approximately constant, $\zeta_4 = 1.64\pm0.03$, and has excellent agreement 
with the value obtained by Benzi {\it et al.} \cite{benzi_jfm_2010} using the extended self-similarity concept~\cite{benzi_PRE_1993}, where $\zeta_4=1.66\pm0.02$. 
It is noteworthy that hyperviscous $\zeta_4$ is precisely inside the interval predicted by the multifractal formalism \cite{benzi_jfm_2010}. 

The remarkable agreement between the Lagrangian statistics from Newtonian and hyperviscous simulations, for inertial time lags, allows us to use 
hyperviscous simulations to study the Lagrangian self-similarity 
in turbulent flows, as discussed below. 

~\\
{\it Self-similarity of the Lagrangian 2nd order structure function}.
%
The new DNS were used to assess whether the 2nd order Lagrangian velocity structure function (LVSF-2) obeys the Lagrangian 
self-similarity relation predicted by Kolmogorov in the form of Eq. (\ref{eq:scaling1941}). The Reynolds numbers attained in the biggest of these simulations is 
$Re_{\lambda} \approx 1700$, which is much higher than previously available \cite{young_pof_2011}. 

In order to prove the Lagrangian self-similarity two conditions have to be fulfilled:
{\it i)} $D^2_L(\tau)$ normalised by $\varepsilon \tau$ must display a plateau, with a universal constant 
$C_0=D^2_L(\tau)/(\varepsilon \tau)$ and, {\it ii)} the same constant must be observed for the three (3) velocity components, since small scale isotropy 
is assumed. Figure \ref{fig:PLOTLVSF_HVISC} shows the LVSF-2 as function of the time lag $\tau$ in logarithmic coordinates,
for several Reynolds numbers/simulations. 
 
As the Reynolds number increases the function $D^2_L(\tau)/(\varepsilon \tau)$ clearly tends to a constant value $C_0 \approx 6.6$.
Specifically, for the simulation with $Re_{\lambda}=1744$ the observed plateau, where $D^2_L(\tau) / \epsilon\tau \ge 0.99 C_0^{\ast}$, is observed for
$3.8 \lesssim \tau/\tau_{\eta} \lesssim 7.0$, which is about 27\% of a decade in $\tau/\tau_{\eta}$.
It is possible to see that after the peak value $D^2_L(\tau) / \epsilon\tau$ at $\tau/\tau_{\eta} \approx 4$, 
there is a region between $5 \lesssim \tau/\tau_{\eta} \lesssim 30-80$ (higher upper limits for higher Reynolds numbers), 
where a new slope, less steeper than $-1$, is observed. It is clear that this secondary slope tends to $\approx 0$ (plateau) as the Reynolds number increases,
an interesting feature that had not yet been observed before.

The second requirement was assessed by analysing the values of the parameter $e=\frac{max(|C_0^*-C_0^i|)}{C_0^*}$. 
Stronger isotropy is recovered for the higher Reynolds numbers cases with $e=0.013,0.011$ and $0.003$, for the simulations
with $Re_{\lambda}=701,1102$ and $1744$, respectively, which shows that isotropy has been recovered in the present simulations, 
and attests that one of the basic assumptions of Kolmogorov's Lagrangian self-similarity is indeed observed here. 
   
\begin{figure}[h]
\centering
\includegraphics[width=0.50\textwidth]{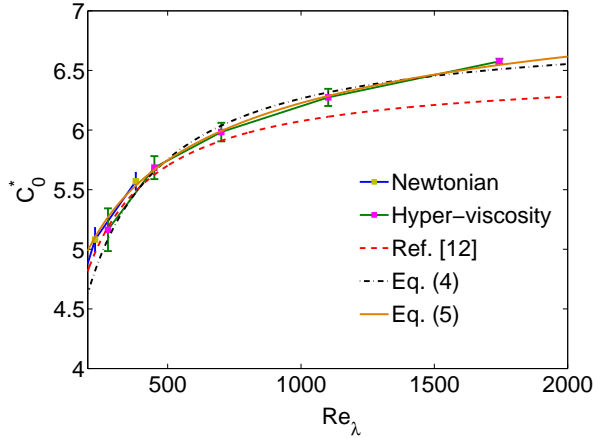}
\caption{
LVSF-2 constant $C_0$, obtained from the new hyperviscous simulations (listed in table \ref{table_DNS}) compared with the empirical relation 
(Eq. \ref{Empi_sawford}) from \cite{Sawford_dispersion}, and new empirical curves obtained with the present new data 
using Eqs. (\ref{Empi_sawford}) and (\ref{Empi_barjona}), respectively. 
}
\label{fig:PLOT_C0_NEWFIT}
\end{figure}

The results from the high Reynolds numbers obtained with the new hyperviscous DNS can be used also to refine the empirical laws  
previously obtained for $C_0^{\ast}$. In reference \cite{Sawford_dispersion} the data available by then was used to determine the coefficients
$a$ and $b$ for a scaling curve with the form,
\begin{equation}\label{Empi_sawford}
C_0^{\ast} = a / \left( 1 + b / Re_{\lambda} \right),
\end{equation}
where in that case the values of $a=6.5$ and $b=70$ were obtained.
%
Considering a similar curve for the present data we obtain $a=6.9$ and $b=1.97$, for an error of $e=0.0026$. 
Figure \ref{fig:PLOT_C0_NEWFIT} shows the comparison of the two curves where one can foresee that 
the present data suggests that the asymptotic value of $C_0^{\ast}$ is clearly higher than previously thought \cite{Sawford_dispersion}.  
We also considered adjusting the present data to a curve of the form,
\begin{equation}\label{Empi_barjona}
C_0^{\ast} = c / \left( 1 + d / Re_{\lambda}^{1/2} \right),
\end{equation}
where constant values of $c=7.8$ and $d=8.0$ are obtained for yet a smaller error, namely $e=0.00092$. 
Given the input from higher Reynolds numbers and the small associated error Eq. (\ref{Empi_barjona}) can be considered 
to be the best approximation for $C_0^{*}$ in existence. 
 
The previous results show that the LVSF-2 nearly exhibits the predicted inertial range plateau to a degree not previously observed, 
however they do not allow one to finally establish the value of the universal constant $C_0$, unless one is prepared to risk something
as crude as extrapolating the data using Eq. (\ref{Empi_barjona}). There are however, a couple of interesting observations that one can 
gather from a close inspection of the LVSF-2, that shed new light on this challenging old problem.

\begin{figure}[h]
\centering
\includegraphics[width=1.00\textwidth]{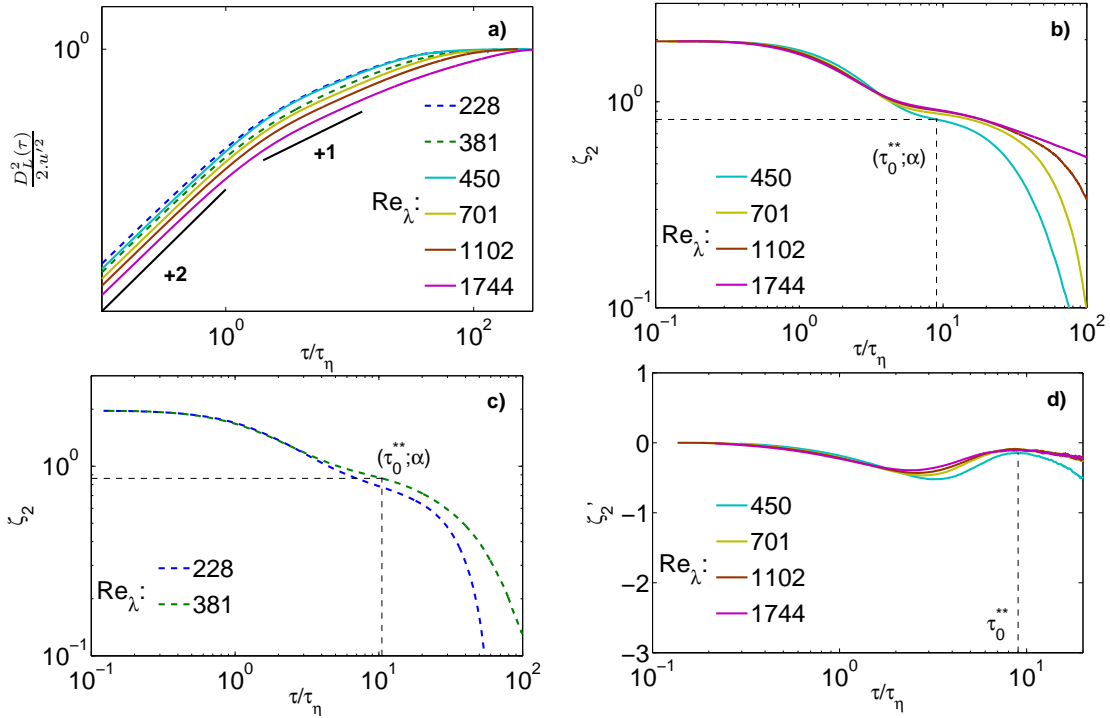}
\caption{
Second order Lagrangian velocity structure function, $D_L^2(\tau)$ (a), and its first $\zeta_2$ (b-hyperviscous, c-Newtonian), and second $\zeta_2'$(d), 
(logarithmic) derivatives, as function of the time lag $\tau$, for all the higher Reynolds simulations used in the present work.   
}
\label{fig:LVSF-2_all}
\end{figure}

To describe these observations Figs. \ref{fig:LVSF-2_all} (a-d) shows the LVSF-2 and its (logarithmic) first and second derivatives defined as,
\begin{equation}\label{eq:primeira_derivada}
\zeta_2 (\tau) = \frac{d(log(D^2_L(\tau)))}{d(log(\tau))},
\end{equation}
and,
\begin{equation}\label{eq:primeira_derivada}
\zeta_2' (\tau) = \frac{d^2(log(D^2_L(\tau)))}{d^2(log(\tau))},
\end{equation}
respectively. 

Already when $D^2_L(\tau)$ is normalised by $2 u'^2$ (Fig. \ref{fig:LVSF-2_all} a) we see the emergence of three different power law regions, associated
with the dissipative ($+2$), inertial ($+1$), and large ($0$) time scales, however we can more clearly observe the emergence of the inertial range plateau 
by analysing the first derivative $\zeta_2 (\tau)$ which is shown in Figs. \ref{fig:LVSF-2_all} (b-hyperviscous, c-Newtonian). 

The curves show a characteristic change of shape around $\tau \approx 10\tau_{\eta} $ for all simulations, and interestingly, the slope of $\zeta_2 (\tau)$ 
following this point tends to decrease as the Reynolds number increases, indicating a tendency for a plateau. Notice that the point where this happens (which we name $\tau_0^{* *}$), 
being one order of magnitude larger than $\tau_{\eta}$, is certainly more likely to carry information regarding the inertial time scales than the 
point near $\tau_0^{*}\approx 4\tau_{\eta}$, typically used to assess $C_0^{*}$, where the viscous effects are still surely felt. 
The inertial range plateau in $D^2_L(\tau)$, if it exists, will be easily observed in $\zeta_2 (\tau)$ through a range of values of $\tau$ where $\zeta_2 (\tau) \approx 1$. 

Finally, the second derivative of $D^2_L(\tau)$, $\zeta_2' (\tau)$, is shown in Fig. \ref{fig:LVSF-2_all} (d) for the hyperviscous simulations. One can see that all the curves 
display a peak near $\zeta_2' (\tau) \approx 0$ at $\tau=\tau_0^{* *}$, and the inertial range plateau in $D^2_L(\tau)$, if it exists, would be observed as a
range of values of $\tau$ where $\zeta_2' (\tau)=0$. We now rigorously define $\tau_0^{* *}$ as the location of this peak in all the simulations. Table \ref{table_DNS}
lists the values of $\tau_0^{* *}$ obtained for the higher Reynolds numbers. Due to some 'noise' in the statistical convergence around $\tau_0^{* *}$
the value of $\tau_0^{* *}$ used here is obtained with a window of width equal to $1\tau_{\eta}$ {\it i.e.} we chose data in a small interval with $[-0.5 \le \tau/\tau_{\eta} \le 0.5]$
centred around $\tau_0^{*}$ to compute this value. 

Using the observations made above we are finally in condition to compute the value of the universal constant $C_0$, directly from our data. 
We define a power $\alpha$ such that one can write, 
\begin{equation}
\label{eq:fundamental}
D^2_L(\tau) = C_0 \epsilon \tau \left( \frac{\tau}{\tau_\eta} \right)^{\alpha-1}.
\end{equation} 
With this definition it follows from Kolmogorov's Lagrangian similarity that for inertial range time lags $\tau_{\eta} \ll \tau \ll \tau_L$ 
and in the asymptotic limit of infinite Reynolds numbers, $\alpha \rightarrow 1$. 
We now note, from the discussion of Figs. \ref{fig:LVSF-2_all} (a-d), that this asymptotic result is concomitant with $\zeta_2 (\tau_0^{* *}) \rightarrow 1$,
so that, in this limit ($Re_{\lambda} \rightarrow \infty$) one can write,   
\begin{equation}
\label{eq:alpha_def}
\alpha = \zeta_2(\tau_0^{* *}),
\end{equation} 
and therefore the universal constant $C_0$ can be computed through,
\begin{equation}
\label{eq:c0**}
C_0^{**} = \frac{ D^2_L(\tau) }{ \left( \epsilon \tau \right) } \left( \frac{\tau}{\tau_\eta} \right)^{1 - \alpha },
\end{equation} 
in the limit of $Re_{\lambda} \rightarrow \infty$ for $\tau_{\eta} \ll \tau \ll \tau_L$. 

\begin{figure}[h]
\centering
\includegraphics[width=0.85\textwidth]{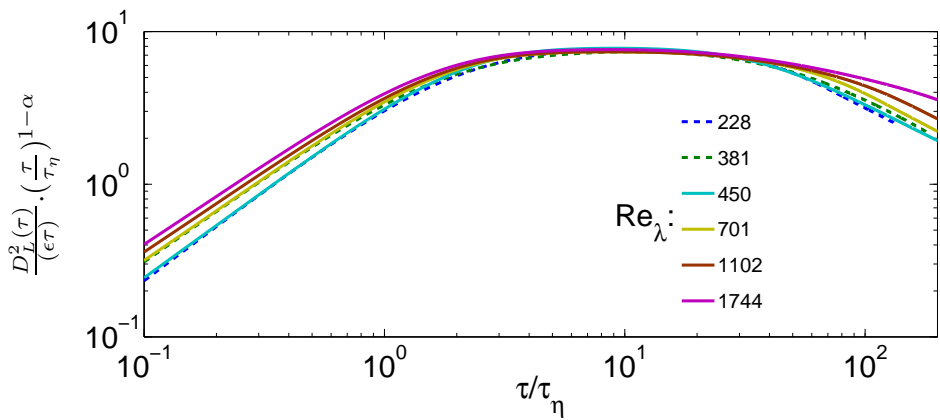}
\caption{
Constant $C_0^{* *}$ obtained from Eq. (\ref{eq:c0**}) for all the simulations used in the present work. 
A constant value of $C_0^{**}=7.5\pm0.2$ is observed for inertial range time lags.
}
\label{fig:descoberta}
\end{figure}

Figure \ref{fig:descoberta} shows $C_0^{**}$ computed with Eq. (\ref{eq:c0**}), for the higher Reynolds simulations used in the present work (Newtonian and hyperviscous). 
It is clear that an inertial range is observed since all the curves collapse for time lags near $\tau_0^{* *}$. Specifically, the width of the inertial range plateau {\it i.e.} 
the interval of time lags in which Eq. (\ref{eq:c0**}) is greater than 99\% of its peak value, is tremendously increased here compared to the plateau associated with 
$C_0^{*} = D^2_L(\tau_0^{*})/(\varepsilon \tau_0^{*})$, which is obvious when comparing Figs. \ref{fig:PLOTLVSF_HVISC} and \ref{fig:descoberta}. 

We now define $\alpha=\zeta_2(\tau_0^{* *})$ for all our (finite Reynolds number) simulations,
thereby extending the definition in Eq. (\ref{eq:alpha_def}). Table \ref{table_DNS} displays the values of $\alpha$ for the higher Reynolds number simulations, 
where the associated possible variation (due to the window used to obtain $\tau_0^{* *}$ described above) is equal to $\pm 0.01$. 
It is noteworthy that $\alpha$ computed using Eq. (\ref{eq:alpha_def}), is precisely the value that maximizes the width of the inertial range plateau. 
Furthermore, we note that as the Reynolds number increases $\tau_0^{*} \rightarrow \tau_0^{* *}$, which further supports the fact that 
$\tau_0^{* *}$ rather than $\tau_0^{*}$ should be used to obtain $C_0$. 

Finally, we compute $C_0^{**}$ through Eq. (\ref{eq:c0**}) and we obtain $C_0^{**} = 7.5 \pm 0.2$ for all the higher Reynolds simulations (see also table \ref{table_DNS}). 
Interestingly, if one uses Eq. (\ref{Empi_barjona}) to estimate the asymptotic value of the Reynolds number that would lead
to $C_0^{\ast}=7.5$ we obtain $Re_{\lambda} \approx 38,000$. This value is not far from the typical value of $Re_{\lambda} \approx 30,000$ 
estimated as the asymptotic Reynolds number needed to obtain $C_0^{*}$ that has been predicted in some references {\it e.g.} \cite{young_pof_2011}. 
Furthermore, we see that the value of $C_0=7.2$ obtained in that paper is also consistent with the present results.

Even though the present results cannot definitely prove Kolmogorov's Lagrangian similarity, in particular regarding the LVSF-2, 
they certainly provide a stronger support for its validity than had been observed thus far, and allow the computation of the universal constant $C_0$
with a new degree of certainty.        

~\\
{\it Conclusions}. 
New Newtonian and hyperviscous direct numerical simulations (DNS) transporting millions of tracers were carried to analyse the second order 
Lagrangian velocity structure function (LVSF-2). The new hyperviscous DNS attain a Reynolds number of $Re_{\lambda} \approx 1700$, which is the highest Reynolds
number attained so far in numerical investigations of Lagrangian turbulence, and it is shown that these hyperviscous simulations can be used to accurately 
compute the LVSF-2 for inertial range time lags. 
The new results shown an unprecedented strong support for Kolmogorov's similarity turbulence theory in a Lagrangian frame, 
and the universal constant defined in the LVSF-2 is computed with a new degree of confidence, giving $C_0 = 7.5 \pm 0.2$.

~\\
{\it Acknowledgements}. We acknowledge PRACE for awarding us access to resource Marenostrum III based in Spain at https://www.bsc.es.
%
%
%


\bibliographystyle{aipnum4-1_pof}

%

\end{document}